\documentclass{article}
\usepackage{times}
\usepackage{boxedminipage}
\usepackage[dvips]{graphicx}

\begin{document}

\title{An Architecture for Security and Privacy in Mobile
        Communications}
\author{G. Winfield Treese \\
  Serissa Research, Inc. \\
  and \\
  Edward R. Murrow Center \\
  Fletcher School of Law and Diplomacy \\
  Tufts University \\
  treese@serissa.com
  \and
  Lawrence C. Stewart \\
  Serissa Research, Inc. \\
  stewart@serissa.com
}

\date{October 1, 2001}

\maketitle

\begin{abstract}
  
  There is much discussion and debate about how to improve the security and
  privacy of mobile communication systems, both voice and data. Most
  proposals attempt to provide incremental improvements to systems that are
  deployed today. Indeed, only incremental improvements are possible, given
  the regulatory, technological, economic, and historical structure of the
  telecommunications system.  In this paper, we conduct a ``thought
  experiment'' to redesign the mobile communications system to provide a
  high level of security and privacy for the users of the system. We discuss
  the important requirements and how a different architecture might
  successfully satisfy them. In doing so, we hope to illuminate the
  possibilities for secure and private systems, as well as explore their
  real limits.

\end{abstract}

\section{Introduction}

As the use of mobile communications devices has increased, many people
have become more concerned with the privacy of such
conversations. Most recently, talk about location-based services has
raised the issue even further, particularly as more people realize
that mobile phones are easily tracked---in fact, the current
architecture of mobile phone systems essentially requires that they be
tracked. 

As the global telecommunications system has evolved, including the
mobile communications system, successive improvements have typically
not been designed with security and privacy in mind. For example,
technological and economic choices by telecommunications providers
often made it possible for governments (and others, in some cases) to
eavesdrop on conversations and obtain call history information.  Over
the past hundred years, in fact, bodies of government laws and
regulations have grown up to institutionalize these abilities, despite
changes in the underlying technology.

The privacy issues become more acute with mobile systems. For example,
it is now much easier for an individual to eavesdrop on cell phones
(although doing so is regulated). Location-based services remind
mobile phone users that they can be tracked. New threats are
appearing, and old threats are becoming more visible.

Discussions about how to improve the privacy of the mobile
communications network typically focus either on regulatory approaches
or layered technologies.  While it is certainly important to consider
and enact policies to protect privacy, such regulations essentially
mean: ``we know you have this information, but you may only disclose
it to specified parties under specified circumstances.'' From a user's
point of view, such protection is, of course, rather limited.

An alternative approach to protecting privacy is to make it difficult,
if not impossible, for anyone to collect the information in the first
place. The means to do this are primarily technological, but not
entirely. Layered technology approaches, such as encrypting phones,
are one approach, but their effectiveness is limited by the underlying
infrastructure. 

In this article, we propose a communications network architecture
designed for security and privacy. The architectural approach is to
separate the network (providing connectivity) from the service
(providing authentication, billing, and so forth). Common carriers
provide the network, while the user can choose among many service
providers based on their pricing, provisions for privacy, and service
offerings.  Security and privacy are provided because the network
cannot know the identity of the users, the service cannot know the
user location, and (by virtue of end-to-end encryption), no one but
the correspondents can know the content of their communications.

An important goal of this work is to explore the limits on how much
security and privacy can be achieved in a large communications
network. Even if the system is unrealistic, or simply impractical to
deploy in place of the existing system, such an exploration can help 
Understanding these limits helps to measure the effectiveness
of other approaches

In order to focus our discussion, we focus mainly on the technologies
and policies in the United States, although the fundamental issues of
both are global in nature. We also focus primarily on mobile voice
communications, although many of these ideas can clearly be applied to
stationary phones and to data networks, whether stationary or mobile. 

\section{Mobile Communications Today}

\begin{figure}[htbp]
  \begin{center}
    \includegraphics[width=\textwidth]{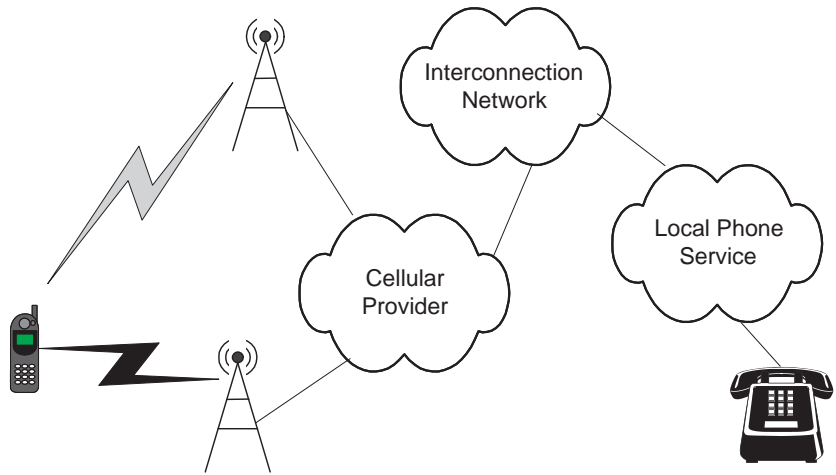}
    \caption{Simplified Architecture of the Cellular Phone System}
    \label{fig:cell}
  \end{center}
\end{figure}

The current wireless networks in North America are based on ``cells'',
which divide a telephone service territory into small regions for
efficient use of low-power transmitters with minimum interference.
The end user devices are generally dumb (indeed, the industry calls
them "terminals") and intelligence---that is, call management,
routing, and other services---is lodged in the network. A simplified
architecture of the cellular network is shown in
Figure~\ref{fig:cell}. A much longer discussion of current
telecommunications systems can be found in
Tomlinson~\cite{Tomlinson00:Telecommunications}.

When a cell phone is turned on, it locates a nearby cell, which tells
it what channels to use and what transmit power to use. As a phone
moves, it will repeat this process as it enters new cells. Phones
identify themselves to the network using a unique ID, called an
\emph{electronic serial number}, or (ESN). The ID is mapped to a
phone number within the network. A common attack on cell phone systems
to obtain fraudulent service is known as ``cloning,'' in which the ESN
is duplicated for another phone. 

For billing, the network where the phone is located sends a billing
record to the phone's service provider. The raw call detail records
are translated according to rates and plans into particular charges
for the end user. The carriers also settle accounts with each other
for the services incurred on the other's network. End users typically
pay a combination of monthly service charge and per-minute charges,
although prepaid plans with only per-minute charges are becoming more
popular. 

The various wireless network technologies in current use have very
different approaches to security. Early analog cell phones have
essentially no security capabilities. The North American digital
standards have voice privacy using encryption, but the system is
notoriously weak. The European GSM standards also support voice
encryption, but they are subject to a variety of straightforward
attacks. 

In the U.S., the Federal Communications Commission (FCC) has required
that location information be available for all mobile phones by
October 31, 2001. While the FCC's regulation was primarily intended to
be used for locating a phone used to call for help in an emergency,
service providers are looking for ways to take advantage of the
information to generate revenue.

\section{Security and Privacy in Mobile Communications}
\label{sec:security}

Every day, telephones are used millions of times for private
discussions, whether business secrets or personal affairs. Most of the
time, the speakers give little thought to the security or privacy of
their communications. What are the issues in a typical call?

First, the parties do not necessarily have a good idea who the other
one is. When speaking to others they know, people can identify them by
voice. Beyond that, there is no real authentication in the system.
Many people easily accept any identification given to them on the
phone: ``I'm from the IRS, and we're just checking up on a problem
with your tax return. Could you give me your Social Security number?''
Similarly, there is no certainty that calling a phone number gets to
the right person or organization, though it is admittedly harder to
persuade someone to call a spoofed number.

Second, someone else may be eavesdropping on the conversation, whether
by a wiretap or by intercepting a conversation broadcast by a cell
phone or cordless phone.\footnote{Note: in this paper we do not
  address the security or privacy of communications between cordless
  phones and their base stations.} Existing standards for voice
privacy on digital (but not analog) phones provide some protection,
but not against a determined attacker. In practice, the technology
discourages simple scanning and opportunistic eavesdropping.

Third, billing records provide an audit trail held by the service
provider of calls made and received, and the phone number (though not
the actual personal identity) of the other party. These records are
often used by law enforcement, and sometimes in other circumstances as
well. 

Fourth, caller ID (CID) and automatic number information (ANI) reveals
the phone number of the caller to the recipient. While we don't
dispute the usefulness of caller ID, it does reveal information that
the caller may wish to keep private. Symmetrically, having a single
phone number used for multiple calls, or by different callers, may
link together information that the recipient may wish to keep private.

Fifth, the weak authentication of devices in the system makes fraud
possible. The mobile telecommunications industry has spent an enormous
amount of money to combat fraud by putting more intelligence in the
network. Here it is important to note a critical distinction about
authentication: devices (e.g., phones) can be authenticated to the
network as legitimate, and users can be authenticated to each other,
regardless of the phones they are using. In fact, it is only important
for the device to be authenticated to the network to link it to a
billing account in the current architecture. We will return to this
problem below.

Finally, it is important to note that solving many of these security
problems requires an \emph{end-to-end} approach. That is, the security
relationship exists between the end users (or, at least, the end
devices), not with parts of the network. For example, consider a
system that encrypts the signal between a wireless phone and the
cellular base station, and then encrypts the signal from the base
station through the phone network. The security is not end-to-end,
because the an intermediate system---the base station---has access to
the clear version. This approach to system design is described in more
detail by Saltzer et al~\cite{Saltzer84:Endtoend}.

These problems of security and privacy are fundamental aspects of the
telecommunications system, and we can measure the effectiveness of a
given system for security by how well it eliminates each threat.

\section{Legal Environment}

Historically, in the United States, security and privacy of mobile
communications has passed through a number of phases. Section 605 of
the Communications Act of 1934 says:

\begin{quotation}
  
  ...no person not being authorized by the sender shall intercept
  any communication and divulge or publish the existence, contents,
  substance, purport, effect, or meaning of such intercepted
  communication to any person...
\end{quotation}

In other words, it was not illegal to listen to a call, but it was
illegal to tell anyone about it.  Scanner and short-wave enthusiasts
had free rein of the airwaves for the next fifty years. In 1984,
however, the Electronic Communications Privacy Act (ECPA) changed the
situation to make it illegal to ``intentionally intercept...any wire,
oral, or electronic communication..'' While the act does provide for
some exceptions, the law essentially made it illegal to listen to
another's conversation without authorization. Later, the
Telecommunications Disclosure \& Dispute Resolution Act of 1992 made
it illegal to manufacture or sell equipment that was capable of
receiving cellular or cordless telephones.


Notwithstanding these U.S. restrictions on the public intercepting
wireless communications, the 1994 Communications Assistance for Law
Enforcement Act (CALEA) requires that carriers implement and as
necessary modify their equipment to facilitate the ability of the
government to intercept communications. An FCC report on the law stated:

\begin{quotation}
  
  Specifically, section 103(a) of CALEA requires that ``a
  telecommunications carrier shall ensure that its equipment,
  facilities, or services that provide a customer or subscriber with
  the ability to originate, terminate, or direct communications'' are
  capable of (1) expeditiously isolating the content of targeted
  communications transmitted by the carrier within its service area;
  (2) expeditiously isolating information identifying the origin and
  destination of targeted communications; (3) transmitting intercepted
  communications and call identifying information to law enforcement
  agencies at locations away from the carrier's premises; and (4)
  carrying out intercepts unobtrusively, so that targets are not made
  aware of the interception, and in a manner that does not compromise
  the privacy and security of other communications.~\cite{FCC99:CALEA}.

\end{quotation}

In a sense, the primary purpose of CALEA was to preserve the ability
of the U.S. government and other law enforcement organizations to
conduct wiretaps and trace telephone calls, an ability that was
jeopardized by evolving technology. Much more background on wiretaps
and communications policy can be found in Diffie and
Landau~\cite{Diffie99:Privacy}.

Separately, the U.S. government has historically discouraged the use
of encryption. While the official policy of the U.S. government is
that no restrictions be placed on the ability of its citizens to use
encryption technology, it has long placed fairly stringent controls on
the export of encryption technology. This has effectively discouraged
the sale of encryption systems in the U.S. market as well. Both the
export regulations and the market conditions for encryption system
have changed in recent years, but such devices as encrypting phones
are still not widely available.

In summary, while it is not illegal for end users to communicate
securely, government policies effectively discourage such systems. In
particular, telecommunications carriers are prohibited from offering
secure services to a broad market, leaving the burden of security
entirely on the end users.

\section{Requirements and Desiderata}

Now we turn to the requirements for a secure and private mobile
communications system. In addition to the basic security requirements
described earlier in section~\ref{sec:security}, there are several
important requirements for the different players in the communications
network. For our simplified architecture, we look at required and
desired properties for four kinds of participants: end users
(including application providers), carriers, and governments.

\subsection{For End Users}

In this system, the category of ``end users'' includes people who make and
receive telephone calls, as well as \emph{application providers}.
Applications providers are those who are providing some service (that is,
an application) over the communications network. We use the term
``application provider'' instead of ``service provider'' to avoid confusion
with those providing telecommunications services (\emph{i.e.},
carriers). The simplest case of an application provider is just a person
answering a telephone call, but other applications may be quite different.

System requirements for end users include:

\begin{enumerate}
\item No one else should be able to bill calls to another account. In
  addition, a stolen phone should be useless, thus discouraging theft.
  
\item The network should keep no record of calls sent or received, but the
  user should have access to complete call detail information. This implies
  that there are no records about uses of digital information services.
  
\item It should not be possible to record a clear version of a
  conversation or data session.
  
\item It should not be possible to discover the location of a user, but the
  user should be able to release her location as desired. For example, a
  stalker with access to the network should not be able to track an
  individual.
  
\item It should not be possible to identify the end user or the end device.
  For example, a device should not have a static ESN.~\footnote{This does
    not rule out ``fingerprinting'' the device based on its transmission
    characteristics, although it is possible to build devices that alter
    their characteristics over time, making fingerprinting difficult, if
    not impossible.}
  
\item Location information is not normally available to anyone, except that
  the network does know the location of a device that is transmitting at a
  particular time. Users can choose to release location information to
  application providers, or it can be automatically released for a call to
  emergency services.
    
\end{enumerate}

In a sense, these are the most important requirements in the system,
because it is the security and privacy of the end users that concerns us
here. Of course, some of them imply a requirement for end-to-end security,
since the network is not trusted.

\subsection{For Carriers}

Telecommunications carriers obviously have many requirements for the
systems that they will build and operate, and they must certainly be
concerned with building scalable, reliable systems. For the architectural
discussion here, we focus on some fundamental requirements that make it
possible to operate the service as a business.

\begin{enumerate}

\item Carriers should be paid for providing services. 

\item The system should have adequate defenses against fraud. 
  
\item There must be mechanisms for naming and addressing end devices, and
  for routing the communications.
  
\item It should be possible to provide add-on features, such as voice mail,
  call forwarding, etc.
\end{enumerate}

We are specifically omitting many aspects of regulated telecommunications
services, such as universal service, regulated pricing, etc., since they
are not directly relevant to this architectural proposal.

\subsection{For Governments}

The primary requirements for governments and telecommunications systems are
as follows:
\begin{enumerate}
\item Provide location information for emergency services
\item Provide access to communications and information about
  communications for law enforcement
\item Provide a robust infrastructure for use in emergencies
\end{enumerate}

Satisfying the first requirement is an important aspect of the
architecture described below. The second is problematic. Certainly
society has an interest in catching criminals, but we strongly oppose
the notion that everyone's privacy must suffer to make law
enforcement's job easier.  We believe that the benefits to society
given by removing security and privacy from the network are outweighed
by the risks of giving the government too much power. The current
executive and judicial branches and their agencies are, we are sure,
staffed only by people with the highest honor and integrity, but it
has not always been so, and may not be again.

\section{An Architectural Proposal}

\begin{figure}[htbp]
  \begin{center}
    \includegraphics[width=\textwidth]{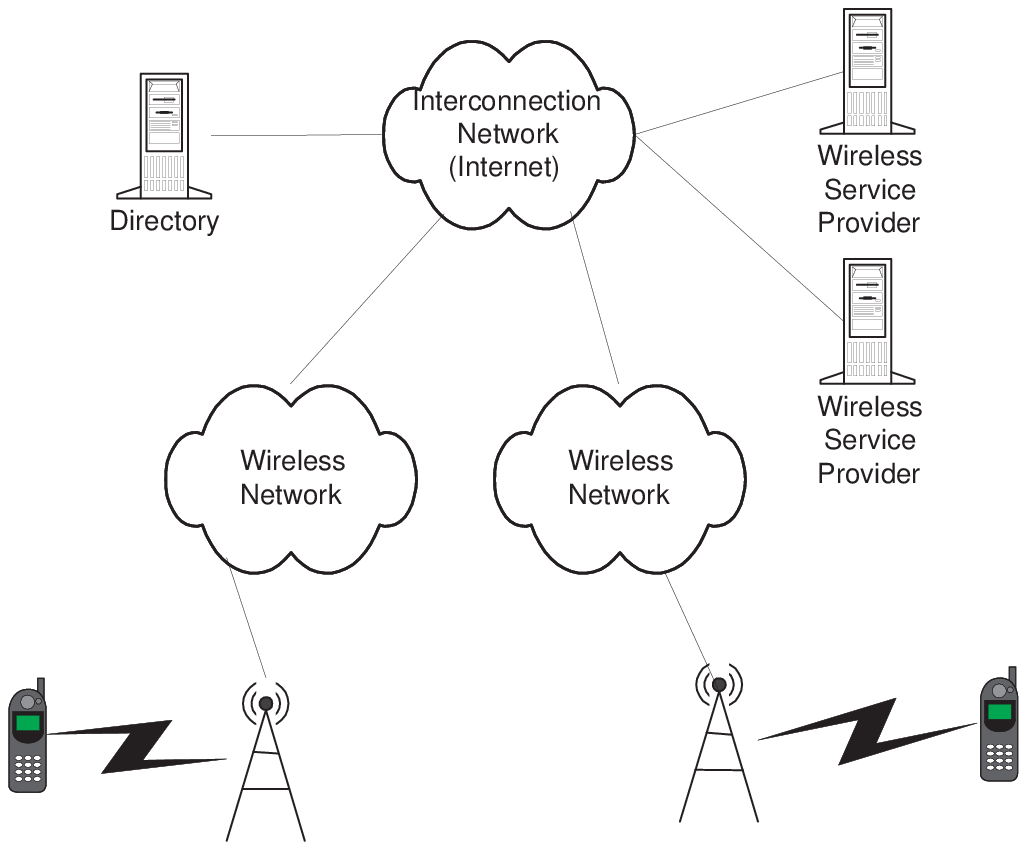}
    \caption{Proposed System Architecture}
    \label{fig:newarch}
  \end{center}
\end{figure}

Our proposed architecture has three main components: the wireless
device (phone), the network, and the service. The role of the network
is to provide the raw communications fabric among wireless devices and
between wireless devices and wireline carriers. The network
\textbf{does not know the identity of the users and does not have a
persistent identifier for the devices}.  The role of the service is to
provide a directory linking the identity of the users to the transient
identifiers of wireless devices and to provide for billing and other
advanced features. The role of the phone is to manage the end-to-end
encryption of communications. A high-level diagram of the system
architecture is shown in Figure~\ref{fig:newarch}.

The easiest way to introduce the architecture is to follow the life
cycle of a call.

When a device (phone) first becomes active, it establishes an
anonymous relationship with the network. In effect the network assigns
a transient ESN, or TESN, to the device. The TESN will be used by the
device for as long as it wants. A privacy-conscious device might
change its TESN for every call, in order to avoid linking the TESN to
an identifiable succession of calls.

Because service has not yet been established, the device cannot make
general connections (except perhaps for emergency calls). However, the
device is allowed a digital connection to the service supplier of its
choice (for convenience, these are identified by URLs). The connection
between the device and the service supplier is made over an end-to-end
encrypted channel, so neither other users nor the network itself can
intercept the communications.

The service provider and the device agree on services and payments.
The service provider will, in turn, pay the network fees for the
activities of the device with the particular TESN.

In order to make a call, the originating device uses the dialed
identifier (the phone number) to look up the service provider of the
destination device in a directory. That service provider is in a
position to supply the destination device's current network provider
and the current TESN within that network. The originating device can
now establish an end-to-end encrypted connection to the destination
device.

\subsection{Naming, addressing, and routing}

In older telephone networks, names, addresses, and routes were
essentially the same: a telephone number. One dialed a particular
number (naming), that number was assigned to a particular phone
somewhere (addressing), and the number included information on how to
route the call (country code, area code, exchange, local identifier).
Human-sensible naming was handled outside the system through paper
directories.

In the modern wireless network, names are represented by telephone
numbers, addresses by ESNs, and routing is managed by the network.
Human-sensible names are still handled outside the system through
paper directories and increasingly through online directories. In
addition, almost every device has a built-in local directory for speed
dialing.

We propose a complete dissociation of naming, addressing, and
routing that is quite analogous to the methods used by Internet
electronic mail (see sidebar).

\begin{figure}[htbp]
  \begin{center}
    \begin{boxedminipage}{\linewidth}

      \begin{center}
        \textbf{Routing Electronic Mail}
      \end{center}
      How does email get from a sender to ``someone@domain.com''?
      Electronic mail on the Internet uses the following iterative
      algorithm:

      \begin{itemize}
      \item The originator uses the Domain Name System (DNS) to locate
        the name of a computer that handles mail for domain.com.
        
      \item The originator then uses the DNS to locate an IP address
        for that computer.
        
      \item The originator transmits the message to the mail server
        for domain.com.
        
      \item The mail server then has two options: it can deliver the
        email to a local mailbox, or it may translate the recipient
        name ``somebody'' into another email address and begins again.
      \end{itemize}
    \end{boxedminipage}
  \end{center}
\end{figure}

\subsubsection{Naming}

Naming will be handled by multiple layers of directories. Devices will
have a search path of directories. The dialed name is given to each
directory in the search path in turn, until a directory either
responds with a translation into a new name, a translation into an
address, or a new directory to try. A directory will typically be
identified by a URL that responds to the directory lookup protocol.
Thus, the ``normal'' path through the name lookup system will
typically require the following steps:

\begin{itemize}
  
\item Look up a name in local device directory (\emph{e.g.,} the speed
  dial directory), which will translate the dialed name ``Mom'' into
  her public \footnote{public in the sense of open to anyone}
  directory entry.
  
\item Look up the public directory entry in the given public directory
  to obtain Mom's service provider and her ID within that provider.
  
\item Look up mom's service-provider-specific-id in the service
  provider's directory to obtain mom's current address.

\end{itemize}

This design provides a number of interesting opportunities:

\begin{itemize}
  
\item A person does not need to place an entry in a public directory.
  Instead, one can place an entry in private directories of one's own
  choosing.
  
\item Directory entries can be have limited access, so that only
  people with appropriate permissions can call you. Indeed, the full
  power of directory systems such as LDAP can be brought to bear, so
  that it is easy, for example, to grant calling permission to groups,
  not just to individuals. Of course, such a system would require
  authentication in order to make a decision about access.
  
\item A person can have many directory entries, in the same or
  different directories. This makes it easy to, for example, to give
  individual names to different callers, which in turn makes it
  possible control, route, or avoid calls from particular callers.
  
\item The lookup process can be delegated to servers on the network,
  so that the process can avoid multiple round trips to wireless
  devices with lower bandwidth and greater latency.
\end{itemize}

By disassociating names in this way, the mechanism becomes both more
flexible and more private, particularly because names can be created
for short-term use and for giving to specific callers for future use.

\subsubsection{Addressing and Routing}

Addresses in this system are anonymous and transient. End devices can
change addresses as often as desired simply by informing their network
(to enable routing and billing) and by informing their service
provider (to enable name translation and billing). An address has the
form $<$network~identifier:TESN$>$, which provides the carrier
network with enough information to handle routing.

Routing will be the province of the networks, and it is their
responsibility to deliver data to the correct device given its
address.

\subsection{Payment}

The end user pays the service provider under whatever terms and using
whatever mechanism they have negotiated, whether that be billing,
credit card, or digital cash. Similarly, the service provider pays the
network using a negotiated mechanism. In this system, it is possible
for service providers to pay in near real-time using a digital cash
system. By doing so, it is possible to have service providers who need
not keep billing records, and therefore leave no information about
what calls were made.

By separating the payment stages in this way, the connection between
the end user and the network is broken: the network cannot identify
the end-user. While the end-user may have limited choice of networks,
he can choose from many alternative service providers, selecting one
based on price, privacy, services offered, or quality of customer
service.

\subsection{Call privacy}

Security of the content of calls will be based on end-to-end
encryption.  For the purposes of the current paper, we will assume
that the encryption algorithm for call content is secure, so that the
primary problem is agreement on a session key. This is more difficult
than it sounds because while two remote devices can easily agree on a
key, using the Diffie-Hellman key exchange algorithm, for example, the
process is subject to man-in-the-middle attacks unless the end devices
have a way to authenticate each other.  Alice is trying to call Bob,
but man-in-the-middle Martin pretends to be Bob to Alice, and pretends
to be Alice, to Bob. In other words, unless we exercise sufficient
care, the caller will get a secure connection, but maybe not to the
correct destination.

There are several approaches to this problem, all of which are
embodied in various other systems today:

\begin{itemize}
  
\item Meet in person. Any two phones, if brought together so that
  their owners can authenticate each other, can agree on keys for use
  in future calls.

\item Public key certificates signed by a certificate authority and
  stored in a directory. A certificate attests to the binding between
  a key and a name.' Because these certificates are signed by a well
  known public key (that of the CA), anyone can check them and
  tampering in the directory is difficult.
  
\item PGP (Pretty Good Privacy) certificates are similar to the public
  key certificates referred to previously, except that there is no
  Certificate Authority. Instead, PGP relies on a ``web of trust''.
  Anyone can sign a key and the software will look for a chain of
  signatures leading to a signature you recognize. If you call the
  same folks frequently, it is easy to build up to high-quality
  authentication.
  
\item Self-signed certificates. A self-signed certificate is a way of
  tying together a name and a key, but anyone can create a self-signed
  certificate for any name they choose and place it in a directory.
  The defense against this attack is for devices to periodically check
  their own entries to assure their accuracy. In case the bad guys
  have subverted the directory to respond correctly only to such
  queries, devices should use a proxy to check the directory entries.
  
\item Anonymous key exchange. While subject to man-in-the-middle
  attacks, this approach prevents eavesdropping attacks.
  Authentication could be accomplished by other means over the secure
  channel.
  
\item SSH approach. In the SSH (``secure shell'') approach, the web
  Secure Session Layer algorithm is used, but with self-signed
  certificates for each end of the connection. There is no special
  attempt to authenticate the initial connection (beyond trusting the
  naming and addressing systems) but since SSH remembers the
  certificates, it is easy to re-authenticate.

\end{itemize}

Ultimately, the end devices themselves (with some help from the end
users) are responsible for the privacy of the communications. Service
providers or the network itself might facilitate authentication,
though the end users would obviously be required to trust them to some
extent.

\subsection{Service Provider}

The service provider fulfills two essential functions:
\begin{itemize}
\item Mapping directory entries to transient network addresses
\item Providing payment to the network carrier
\end{itemize}

The service provider may also provide value added services such as
\begin{itemize}
\item Providing a customer statement with call detail
\item Forwarding calls for privacy
\item Voice mail
\end{itemize}

Let us consider two examples of service providers, occupying opposite ends
of the privacy spectrum.

\subsubsection{Traditional Service Provider}

The traditional service provider makes wireless service under the new
architecture look much like current service. The phone is given a more
or less permanent number, and the service provides a directory which
translates that number into the phone's current transient ESN. The
service provider accepts billing records from network carriers for the
phone's activities, pays the carriers, and provides a traditional
detailed billing statement for the end user. The identity of the
end-user is well known.

\subsubsection{High-Privacy Service Provider}

The high-privacy service provider goes to extreme lengths to safeguard
the privacy of the end-user. This service provider is much like a
current phone card operator, except that the user can receive calls as
well as place them.

\begin{itemize}
\item The service provider does not know the end-user identity,
  instead having a (digital) cash relationship with the customer.
\item The provider creates, as requested, any number of single-use ids
  which map to the phone's transient ESN. The end-user can give out a
  different ``phone number'' to every correspondent.
\item The service provider can operate a relay service, so that calls
  are relayed through a static location, thus concealing from the
  remote end even the transient ESN. The relay does not have access to
  call content - that is still end-to-end encrypted.
\item The service provider can pay network carriers with cash, and
  consequently have no need to retain billing records.\ 
\item The service provider does not provide statements.
\end{itemize}

Between these two extremes is a wide variety of opportunities for
service providers to stake out markets based on price, payment
options, privacy of customer records, etc.

\section{Can we get there from here?}

Is this a realistic system? Is it one that could be deployed as an
evolution of the current infrastructure? From a technical point of
view, we think the answer is yes. The major changes needed to
implement this architecture are in the billing and directory systems,
and in the deployment of devices that are capable of secure
communications. Because wireless networks are already capable of
managing call routing to ESNs that suddenly appear on their network,
the core infrastructure does not change much. It should also be
possible to phase in many of the changes proposed here, so it is not
necessary to have a brand-new system deployed at once.

The economic case is somewhat harder. In part, this is because the
architecture depends somewhat on separating some functions---the
service provider and the carrier network---that are often combined
today (albeit ones that are interconnected with other carriers). There
is no obvious incentive for companies to unbundle their current
services to implement the proposed architecture. A second problem is
that changing the system to improve security would likely cost more
for customers, and it is not clear that customers would value the
security and privacy improvements enough to pay for them.

Finally, as we have noted, the legal and regulatory systems, in the
U.S. as well as other countries, discourage secure communications. In
fact, carriers implementing the proposals described here might
possibly be in violation of CALEA. 

\section{Conclusions}

Today's mobile communication system provides few assurances about the
security and privacy of one's communications. In this paper, we have
proposed a high-level system architecture for a communications system
that does provide strong guarantees for security and privacy for end
users. We believe that the architecture is technologically feasible to
deploy on a large scale, but the current market and legal environments
will certainly not encourage its use.

This proposal demonstrates, however, that it is possible to have such
strong guarantees. Therefore, when technologists, policy makers, and
businesspeople are evaluating options to increase security and privacy
in the telecommunications network, they can use the strong guarantees
as a benchmark for comparison in how effective the options are.

\bibliographystyle{plain}
\bibliography{wireless}

\begin{thebibliography}{1}

\bibitem{FCC99:CALEA}
Federal~Communications Commission.
\newblock 99-229, {Second Report and Order}, August 26 1999.

\bibitem{Diffie99:Privacy}
Whitfield Diffie and Susan Landau.
\newblock {\em Privacy on the Line: The Politics of Wiretapping and
  Encryption}.
\newblock MIT Press, 1999.

\bibitem{Saltzer84:Endtoend}
Jerome~H. Saltzer, David~P. Reed, and David~D. Clark.
\newblock End-to-end arguments in system design.
\newblock {\em ACM Transactions on Computer Systems}, 2(4):277--288, 1984.

\bibitem{Tomlinson00:Telecommunications}
Clive Tomlinson.
\newblock {\em Telecommunications: An Introduction for Software Professionals}.
\newblock Addison-Wesley Publishing Company, 2000.

\end{thebibliography}

\end{document}